\documentclass[12pt]{article}
\usepackage{amsmath,amsthm,amscd,amsfonts,amssymb}
\usepackage{latexsym}
\newcommand{\CC}{\mathbb C}
\newcommand{\BB}{\mathcal B}
\newcommand{\LL}{\mathcal L}

\newcommand{\PP}{\mathbb P}
\newcommand{\RR}{\mathbb R}

\newcommand{\ZZ}{\mathbb Z}
\newcommand{\EE}{\mathbb E}
\newcommand{\oo}{\mathcal O}

\newcommand{\hh}{\mathcal H}

\newtheorem{thm}{Theorem}[section]
\newtheorem{lemma}[thm]{Lemma}
\newtheorem{cor}[thm]{Corollary}
\newtheorem{prop}[thm]{Proposition}
\newtheorem{defin}[thm]{Definition}
\newtheorem{rem}[thm]{Remark}
\newtheorem{hyp}[thm]{Hypothesis}

\def\half{\frac{1}{2}}
\def\pf{{\noindent \bf Proof: }}
\begin{document}
\author{M Krishna \\ Institute for Mathematical Sciences \\ Taramani Chennai 600 113 \\ India
\\
{\small \it Dedicated to Michael Demuth on his 60$^{th}$ birthday}
}
\title{Continuity of integrated density of states -- independent randomness}
\maketitle
\abstract{In this paper we discuss the continuity properties of the integrated
density of states for random models based on that of the single site 
distribution.
Our results are valid for models with independent randomness with arbitrary
free parts.
In particular in the case of the Anderson type models (with stationary, growing, decaying randomness) on the $\nu$ dimensional 
lattice, with or without periodic and almost periodic backgrounds,  
we show that if the single site distribution is uniformly 
$\alpha$-H\"older continuous, $ 0 < \alpha \leq 1$, then
the density of states is also uniformly $\alpha$-H\"older continuous. 
}

\section{Introduction}
In the spectral theory of random potentials one of the quantities of interest
is the density of states which is an averaged total spectral measure.  Often
this measure is approximated using a sequence of operators (finite volume operators) and the continuity properties of the limit are proved using the total
spectral measures of the approximants.  One of the main questions in such
an approximation procedure is the
existence of the limiting density of states measure.

One of the first results on the $\alpha$-H\"older continuity of the integrated
 density 
of states in the case
of singular single site distributions (done for the approximating operators)) is by Carmona-Klein-Martinelli \cite{ckm}, (Lemma 6.1 and Theorem 6.2).
They have essentially done what we present below, but because they restrict
themselves to the approximants of the random operators restricted to boxes,
they do not obtain the generality presented here.  

The literature on the existence of the density of states and the Wegner estimate(Wegner \cite{weg}) 
is vast and we refer to the books Cycone-Froese-Kirsch-Simon \cite{cfks}, 
Carmona-Lacroix \cite{cl}, Figotin-Pastur \cite{fp} for a historical development
of the study of density of states and some of the recent
papers of Combes-Hislop-Klopp-Nakamura \cite{chkn}, Kirsch-Veselic \cite{kv} and Hundertmark-Killip-Nakamura-Stollmann-Veselic \cite{hknsv} for complete
references for more recent advances.  The latest review paper of Werner Kirsch
and Bernd Metzger \cite{km} is a good starting point. 

In this paper we consider a direct proof without going through the approximation
process and this requires us to declare some average total spectral measure
as the density of states and we will choose a definition as in Krishna \cite{mk1} that agrees with the standard one in the case of the Anderson model.

To this end let $\hh$ be a separable Hilbert space and let $(\Omega, \BB_\Omega, \PP)$ be a probability space.  We consider a self-adjoint operator
valued random variable $A$ and a real valued random variable $q$ on $\Omega$.
Thus for each $\omega$, $A^\omega$ is a self-adjoint operator and the 
resolvents $(A^\omega +i)^{-1}$ are weakly measurable in $\omega$ 
(and hence so are $(A^\omega - z)^{-1}, ~ Im(z) \neq 0$).  
$q^\omega$ is a measurable real valued function.  

\begin{hyp}
\label{hyp1}
Let $(\Omega, \BB_\Omega, \PP)$ and $A, q$ be as above.
We shall assume that
$q$ and $A$ are independent, which means that for any vectors $f, g \in \hh$,
the random variables $q$ and $\langle f, A g\rangle$ are independent and
so are $\langle f, \psi(A) g\rangle$ for any bounded measurable function $\psi$
on $\RR$.
\end{hyp}

\begin{defin}
\label{def1}
Let $\mu$ be a probability measure on $\RR$.  Then $\mu$ is said
to be uniformly $\alpha$-H\"older continuous,
$0 < \alpha \leq 1$, if 
$$
\sup_{x \in \RR} ~ ~ ~ \sup_{ 0 < \epsilon \leq 1 } \frac{\mu( ( x-\epsilon, x+\epsilon) )}{(2\epsilon)^\alpha}  < \infty.
$$
Note that the above condition is equivalent to 
$$
d_\mu^{\alpha, \infty} = \sup_{x \in \RR} \sup_{\epsilon \in (0, \infty)} \frac{\mu((x-\epsilon, x + \epsilon))}{(2\epsilon)^\alpha} < \infty. 
$$
\end{defin}

In the definition above we formulated the $\alpha$-H\"older continuity of 
a measure, however this implies the same for the distribution function
of the measure and so often one can say ``the density of states'' is uniformly
$\alpha$-H\"older continuous`` or ''the integrated density of states is uniformly $\alpha$-H\"older continuous`` interchangeably.

Let us consider a pair of operators $A, q$ as in hypothesis \ref{hyp1}.  Let 
$\EE$ denote taking averages of complex valued functions of $A$ with respect
to the measure $\PP$. Thus if $f : \LL(\hh) \rightarrow \CC$, is a bounded
measurable function, then 
$$
\EE (f(A))  = \int f(A^\omega) d\PP(\omega).
$$
Let $q$ be distributed according to the probability measure $\mu$.

Then we have the following theorem, where we denote by $E_B(\cdot)$, the
(projection valued) spectral measure of the self-adjoint operator $B$.
We also denote by $P_\phi$ the orthogonal projection onto the one dimensional
subspace generated by the vector $\phi$.  In what follows the constant
$d_\mu^{\alpha, \infty}$ associated with a measure $\mu$ is given in
Definition \ref{def1}.

\begin{thm}
\label{thm1.1}
Let $A, q$ be as in hypothesis \ref{hyp1} and let
$\phi$ be a unit vector in $\hh$.  Consider
the operators 
$$
H^\omega = A^\omega + q^\omega P_\phi
$$
and consider the measures
$$\nu_{_{A^\omega}} = \int \langle \phi, E_{H^\omega} (\cdot) \phi \rangle ~ d\mu(q^\omega). $$
Suppose $q$ is distributed according
to a probability measure $\mu$ which is uniformly $\alpha$-H\"older 
continuous, $0 < \alpha \leq 1$.
\begin{enumerate}
\item Then $\nu_{_{A^\omega}}$ is uniformly $\alpha$-H\"older continuous with the same exponent $\alpha$ for each fixed $A^\omega$.  
\item We have the following uniform bound for each $\omega$.
$$
d_{\nu_{_{A^\omega}}}^{\alpha, \infty} \leq 2^{2-\alpha} \pi ~ d_\mu^{\alpha, \infty}.
$$
\item $\EE \left( \nu_{_{A^\omega}} \right)$ is also uniformly $\alpha$-H\"older continuous.
\end{enumerate}
\end{thm}

Strictly speaking one should state the theorem with $H = A + qP_\phi$, 
and use the notation $\nu_A$ etc., but we follow the spectral theory 
community's convention of writing 
$H^\omega$ etc., to distinguish the ``random'' operators from the ``deterministic'' operators. 

\begin{rem}\label{rem0.0}
In the above theorem instead of uniform $\alpha$-H\"older continuity we could
also take some modulus of continuity 
$$
s(\mu, \epsilon) = sup_{\{I : |I| < \epsilon\}} \mu(I)
$$
the sup taken over intervals $I$, then the theorem is true for such
a modulus of continuity. This remark is private communication by
Peter Stollmann \cite{sto}.
\end{rem}

As an application of the above theorem we have the following.  Consider
$\Omega = \RR^{\ZZ^d}, \PP = \prod \mu$, 
$\hh = \ell^2(\ZZ^\nu)$ and consider the models 
\begin{equation}
\label{eqn1.1}
H^\omega = \Delta + B V^\omega
\end{equation}
with $(\Delta u)(n) = \sum_{|i|=1} u(n + i)$, $(V^\omega u)(n) = \omega(n) u(n)$, so $\{V(n)\}$ are real valued i.i.d random variables. 
Here $B$ is a real valued  diagonal operator 
$(B u)(n) = a_n u(n)$, with the sequence $\{a_n\}$ being non-zero.
In the case when $B = I$ the identity operator, one has the Anderson model 
which is stationary.

\begin{rem}
\label{rem1.1}
We note that if $\mu$ is uniformly $\alpha$-H\"older continuous and if $c$
is a non-zero real number, then the measures $\mu_c, \mu^c$ defined by
$$
\mu_c(B) = \mu(c B), ~ \mathrm{for \ all} ~ B \in \BB_\RR  ~ \mathrm{or} ~ 
\mu^c(B) = \mu(B + c ), ~ \mathrm{for \ all} ~ B \in \BB_\RR, 
$$
are also uniformly $\alpha$-H\"older continuous.  We note also that by
an easy calculation one has for $c \neq 0$, 
\begin{equation}
\label{eqnn1.1}
d_{\mu_c}^{\alpha, \infty} = |c|^\alpha d_\mu^{\alpha, \infty} ~ \mathrm{and} ~ d_{\mu^c}^{\alpha, \infty} = d_\mu^{\alpha,\infty}.
\end{equation}
\end{rem}

The following theorem is then a corollary of theorem \ref{thm1.1}, which is seen
by setting, $\Omega = \RR^{\ZZ^\nu}$, $\PP = \prod \mu$,  
$\phi = \delta_n$, $V^\omega$ is multiplication by $\omega(n)$ ($\omega$ coming
from the support of $\PP$) and 
$A^\omega = \Delta + BV^\omega - a_n \omega(n) P_{\delta_n}$ with 
$q^\omega = a_n \omega(n)$ for each $n \in \ZZ^\nu$.  Therefore
$q^\omega(n)$ are distributed
according to $\mu_{a_n^{-1}}$ and  we have the following theorem
whose proof mimics the proof of theorem \ref{thm1.1} using the above
facts in the last few steps.

\begin{thm}
\label{thm1.2}
Consider the self-adjoint operators $H^\omega$ given in equation 
(\ref{eqn1.1}). Suppose $V(n)$ are distributed according to a probability
measure $\mu$.
Let $\nu_n$ denote
the measure 
$$
\nu_n(\cdot) = \EE\left(\langle \delta_n, E_{H^\omega}(\cdot)  \delta_n \rangle \right).
$$
Suppose $\mu$ is  uniformly $\alpha$-H\"older continuous 
for some $0 < \alpha \leq 1$,  
then $\nu_n$ is also uniformly $\alpha$-H\"older continuous with exponent $\alpha$.
Any total spectral measure $\nu = \sum_{n} \beta_n \nu_n, ~ \beta_n > 0, ~  \sum \beta_n  = 1$ such that $\sum \beta_n |a_n|^{-\alpha} < \infty$ 
is also uniformly
$\alpha$-H\"older continuous.
\end{thm}

In the case when $B = I$, the above model in equation (\ref{eqn1.1})  reduces to the Anderson model and 
all the measures $\nu_n$ are the same and agree with the density of states
of the Anderson model.  Therefore we have the following corollary.

\begin{cor}
\label{cor1}
Consider the Anderson model $H^\omega = \Delta + V^\omega$, on 
$\ell^2(\ZZ^\nu)$, with $V(n)$ i.i.d distributed according to $\mu$.  
Suppose the stationary distribution $\mu$ is uniformly $\alpha$-H\"older continuous with exponent 
$\alpha$, $0 < \alpha \leq 1$, then the density of states is
also uniformly $\alpha$-H\"oder continuous with the same exponent $\alpha$.
\end{cor}

\begin{rem}
\label{rem1.2}
\begin{itemize}
\item In the equation (\ref{eqn1.1}), we could have replaced $\Delta$ with any
self-adjoint operator.  Thus $\Delta$ perturbed by a periodic perturbation 
is covered. In fact we can take any orbit $\oo_S$ of a  
subset $S$ of $\Omega$ under the $\ZZ^d$ action (by translation) and take a 
nice probability measure on this orbit.  Then
if we take any real valued random variable $W$ supported by $\oo$  and take the operators $\Delta + W + V$, the theorem is still valid when we average over all the randomness $W$ and $V$.  Thus the above theorems cover periodic and almost
periodic backgrounds.
\item Since theorem \ref{thm1.1} is quite abstract it can be phrased in terms
of ergodic and non-ergodic dynamical systems and gives numerous corollaries 
for average spectral measures of the associated self-adjoint operators.  
\end{itemize}
\end{rem}

Finally we mention that in a forthcoming paper with Werner Kirsch \cite{kk} 
we will
consider models of the form  
$$
-\Delta + W + \sum_{i \in \ZZ^d} q_i \chi_{\Lambda_i}, 
$$
on $L^2(\RR^d)$ where $\Lambda_i$ are cubes centred at $i \in \ZZ^d$.
We show that a class of averaged total spectral measures
have the same continuity properties as the single site distribution $q_i$
provided $q_i$ are independent.  Here again the cases cover periodic 
backgrounds and other free parts.

After this work was done, we came to know about the paper of 
Combes-Hislop-Klopp \cite{chk} on the Wegner estimate for the continuous models,
however our work is done independently.  

\section{Proofs}

We begin with a Lemma on Borel transforms, where given a probability
measure $\sigma$ we denote $F_\sigma(z) = \int \frac{1}{x-z} ~ d\sigma(x)$.

\begin{lemma}
\label{lem2.1}
Let $\sigma$ be a probability measure on $\RR$.  Then for any $y \in \RR$ and
any $a \in \RR \setminus \{0\}$ we
have the uniform bound
$$
|\frac{a}{Im (F^{-1}_\sigma(y + ia ))} |  \leq 2 .
$$
\end{lemma}
\pf We have
\begin{equation}
\label{eqn2.0}
\begin{split}
\frac{a}{Im (F^{-1}_\sigma(y + ia ))} &=  \frac{1}{Im ((aF_\sigma)^{-1}(y + ia) )} \\
&= - \frac{ [Re( aF_\sigma(y+ia))]^2 + [Im ( aF_\sigma(y+ia))]^2}{ Im (a F_\sigma(y+ia))}, 
\end{split}
\end{equation}
using the fact that 
$$
Im (z^{-1}) = \frac{-Im(z)}{ (Re(z))^2 + (Im(z))^2)}.
$$
Now we have
$$
aF_\sigma(y+ia) = \int \frac{a}{x - y -i a} ~ d\sigma(x) = 
\int \frac{1}{\frac{x - y}{a} -i } ~ d\sigma(x). 
$$
If we set $\frac{x -y}{a} = \beta(x, y, a)$, then $\beta(x,y, a)$ is real
valued and we have
$$
aF_\sigma(y+ia) = \int \frac{1}{\beta(x, y, a) -i } ~ d\sigma(x).
$$
Using this relation and computing the real and imaginary parts of $aF_\sigma(y+ia)$ we have
\begin{equation}
\begin{split}
\label{eqn2.1}
Re(aF_\sigma)(y+ia)) &= \int \frac{\beta(x,y,a)}{\beta(x,y,a)^2 + 1} ~ d\sigma(x) \\ | Re(aF_\sigma)(y+ia))| &\leq \int \frac{1}{\sqrt{(\beta(x,y,a)^2 + 1)}} ~ d\sigma(x)\\
Im(aF_\sigma)(y+ia)) &=  \int \frac{1}{\beta(x,y,a)^2 + 1} ~ d\sigma(x) \\
\end{split}
\end{equation}
Using these two inequalities and the fact that $\sigma$ is a probability
measure we see that 
\begin{equation}
\label{eqn2.2}
|Im(aF_\sigma)(y+ia))| = \int \frac{1}{\beta(x,y,a)^2 + 1} ~ d\sigma(x) 
\leq 1.
\end{equation}
By using the inequality (\ref{eqn2.1}) and the Schwarz inequality and the inequality (\ref{eqn2.2}), we also have 
\begin{equation}
\label{eqn2.3}
\begin{split}
| Re(aF_\sigma)(y+ia))|  &\leq \int \frac{1}{\sqrt{(\beta(x,y,a)^2 + 1)}} ~ d\sigma(x)
\\ & \leq \left( \int \frac{1}{\beta(x,y,a)^2 + 1} ~ d\sigma(x) \right)^\half \\
&= |Im (a F_\sigma(y+ia))|^\half. 
\end{split}
\end{equation}
Therefore we immediately get the bounds
\begin{equation}
\label{eqn2.4}
\frac{\left(Re(aF_\sigma(y+ia))\right)^2}{|Im(a F_\sigma(y+ia))|} \leq 1 ~ ~ \mathrm{and} ~ ~ \frac{\left(Im(aF_\sigma(y+ia))\right)^2}{|Im(a F_\sigma(y+ia))|} \leq 1.
\end{equation}
This estimate together with equation (\ref{eqn2.0}) gives the lemma. \qed

In the following proposition we give an equivalent condition, in terms of
a wavelet transform of the probability measure $\mu$, for it to be
a uniformly $\alpha$-H\"older continuous measure.

\begin{prop}
\label{pro1.0}
Suppose $\psi$ is a continuously differentiable positive even function on $\RR$ 
satisfying 
$|\psi(x)| + (1+|x|)|\psi^\prime (x)|$ is integrable and $\psi(0) = 1$.
Suppose $\mu$ is a probability measure on $\RR$.  Then, 
for each $0 < \alpha \leq 1$, 
$$
\sup_{x\in \RR}\sup_{a >0}\frac{1}{a^\alpha}(\psi_a*\mu)(x) < \infty  ~ \iff ~ d_\mu^{\alpha, \infty} < \infty. 
$$
\end{prop}
\pf 
The lemma is proved if we show that $\frac{1}{a^\alpha}(\psi_a * \mu)(x)$ 
to be uniformly bounded in $x, a$ if and only if $\mu$ is $\alpha$-H\"older continuous.

To see the if part of this statement, we note the relation 
\begin{equation}\label{eqn000}
\frac{1}{a^\alpha}(\psi_a * \mu)(x) = - \int_0^\infty \psi^\prime(y)(2y)^\alpha \frac{\mu((x-ay, x+ay))}{(2ay)^\alpha} ~ dy,
\end{equation}
as in equation (1.3.4) of Demuth-Krishna \cite{dk}.  
We note that for any $x\in \RR$ and any $a, y \in (0, \infty)$, 
\begin{equation}
\label{eqn001}
\begin{split}
\frac{\mu( (x-ay, x+ay))}{(2ay)^\alpha} &\leq \sup_{ay > 0 } \frac{\mu( (x-ay, x+ay))}{(2ay)^\alpha} \\ 
& \leq \sup_{x \in \RR}  \sup_{ay > 0 } \frac{\mu( (x-ay, x+ay))}{(2ay)^\alpha} \\ & = d(\mu, \alpha) < \infty,
\end{split}
\end{equation}
for some constant $d(\mu, \alpha)$, by the uniform $\alpha$-H\"older continuity of $\mu$ 
(see definition \ref{def1})).
Therefore 
the right hand side is uniformly bounded in $x, a$ since 
$|-\psi^\prime(y)(2y)^\alpha| \leq 2|(1+|y|)\psi^\prime(y)|$ is an 
integrable function, $0 < \alpha \leq 1$ showing that
$\frac{1}{a^\alpha} (\psi_a * \mu)(x)$
is uniformly bounded in $x, a$.

To see the only if  part of the statement, note that since $\psi$ is positive
and continuous with $\psi(0) = 1$, there is a $\beta >0$ depending only on
$\psi$ such that 
$\psi(x) \geq \half, ~ x \in (-\beta, \beta)$.  
So we have
\begin{equation}
\label{eqn002}
\frac{1}{a^\alpha} \psi_a *\mu(x) \geq \frac{1}{a^\alpha} \int_{x-\beta a}^{x+\beta a} \psi_a(y-x) d\mu(y) \geq \half \frac{1}{a^\alpha} \mu((x-\beta a , x+\beta a)). 
\end{equation}
Since $\beta$ is a fixed positive number, it is easy to see that the $\mu$ is uniformly $\alpha$-H\"older continuous
whenever the left hand side of the above inequality is uniformly bounded in $(x, a)$.
\qed

We have a corollary of the above for Borel transforms.  Recall that
$F_\sigma(z) = \int \frac{1}{x-z} ~ d\sigma(x).$
\begin{lemma}
\label{lem2.2}
Suppose $\mu$ is a probability measure on $\RR$.
Let, $0 < \alpha \leq 1$, then 
$$
\sup_{z : Im(z) \neq 0}  ||Im(z)|^{1-\alpha} Im(F_\mu(z))| < \infty ~ \iff ~ d_{\mu}^{\alpha, \infty} < \infty.
$$
In addition we have the bound, 
\begin{equation}
\label{thebound}
\sup_{z : Im(z) \neq 0}  ||Im(z)|^{1-\alpha} Im(F_\mu(z)) | \leq 2^\alpha \pi d_\mu^{\alpha, \infty}.
\end{equation}
\end{lemma}
\pf We set $\psi(x) = \frac{1}{1+x^2}$ in which case the first part of the
lemma is valid by setting $z = E+ia$ so that
$$
Im(F_\mu(z)) = \frac{1}{a}(\psi_a * \mu)(x), ~ \mathrm{and} ~ ~ |Im(z)|^{1-\alpha} F_\mu(z) = 
\frac{1}{a^\alpha} (\psi_a * \mu)(x), 
$$
where we have taken $\phi * \mu (x) = \int \phi(y-x) d\mu(y)$.
Hence the result follows for the case of $a >0$ from Proposition \ref{pro1.0}. 
From the equation \ref{eqn000} and the inequality  (\ref{eqn001}) we see that, 
$$
|\frac{1}{a^\alpha} (\psi_a*\mu)(x) | \leq d_\mu^{\alpha, \infty} \int_0^\infty |\psi^\prime(y) (2y)^\alpha| ~ dy, 
$$
which gives the bound 
$$
\sup_{x\in\RR} \sup_{a >0} |\frac{1}{a^\alpha} (\psi_a*\mu)(x) | \leq d_\mu^{\alpha, \infty} 2^\alpha \pi,
$$
by making use of the fact that $\psi(x) = \frac{1}{1+x^2}$ in the present case.
Since $\psi$ is even,  $\psi_{-a}(x) = \psi_{a}(x)$,  so that the lemma 
for $Im(z) < 0 $ follows from that for $Im(z) >0$. \qed 

\begin{lemma}
\label{lem2.3}
Let $\sigma$ be a probability measure and let $\mu$ be a probability measure
which is uniformly $\alpha$-H\"older continuous, $0 < \alpha \leq 1$, 
and let $d_\mu^{\alpha, \infty}$ be the constant given in the 
Definition \ref{def1}. 
Then we have 
\begin{equation}
\label{eqn2.5}
\sup_{y \in \RR } ~ \sup_{ a > 0} | a^{1-\alpha} Im \left( \int \frac{1}{ x + F_\sigma(y+ia)^{-1}} ~ d\mu(x)\right) | < 2\pi ~  d_\mu^{\alpha, \infty}.
\end{equation}
\end{lemma}
\pf We first note that since $\sigma$ is a probability measure on $\RR$,
the function $F_\sigma(z)$ has non-zero imaginary part whenever $z$ has
non-zero imaginary part.  Therefore we have
\begin{equation}
\label{eqn2.6}
\begin{split}
& \left|a^{1-\alpha} Im \left( \int \frac{1}{ x + F_\sigma(y+ia)^{-1}} ~ d\mu(x)\right) \right| 
\\ &= 
\left|\left(\frac{a}{Im(F_\sigma^{-1}(y+ia))}\right)^{1-\alpha} \right| \\
& ~ ~ \times \left| (Im(F_\sigma^{-1}(y+ia)))^{1-\alpha} Im \left( \int \frac{1}{ x + F_\sigma(y+ia)^{-1}} ~ d\mu(x)\right) \right|\\
& \leq 2^{1-\alpha} 2^{\alpha} \pi ~ d_\mu^{\alpha, \infty}, 
\end{split}
\end{equation}
where Lemma \ref{lem2.1} gives the bound $2^{1-\alpha}$ for the first factor
while the second factor is bounded by 
$\sup_{w : Im(w) \neq 0} |Im(w)|^{1-\alpha} |Im(F_\mu(-w)|$, 
by taking $w = F_\sigma(y+ia)^{-1}$, 
which is bound using Lemma \ref{lem2.2},  in view of the uniform 
$\alpha$-H\"older continuity of $\mu$. \qed  

{\noindent \bf Proof of Theorem \ref{thm1.1} :}  
The parts (1) and (3) are obvious if we prove (2), so we restrict ourselves to
proving (2). 
Let us define

\begin{equation}\label{eqn2.7}
\begin{split}
F^\omega(E+ia) &= \langle \phi, (A^\omega + q^\omega P_\phi - E - ia )^{-1} \phi \rangle, 
~ \mathrm{and} ~  \\ 
F^\omega_0(E+ia) &= \langle \phi, (A^\omega - E - ia )^{-1} \phi \rangle. 
\end{split}
\end{equation}
Using the spectral theorem we have
$$
\int \frac{1}{x - E - ia} ~ d\langle \phi, E_{A^\omega + q^\omega P_\phi}(x) \phi \rangle = F^\omega(E+ia).
$$
Then we have, taking average of $q^\omega$ with respect to $\mu$, using the
definition of $\nu_{A^\omega}$ and using Fubini, 
\begin{equation}\label{eqn2.71}
\int \frac{1}{x - E - ia} ~ d\nu_{_{A^\omega}}(x) = \int F^\omega(E+ia) ~ d\mu(q^\omega) 
\end{equation}
and from Lemma \ref{lem2.2} it is enough to show that 
\begin{equation}\label{eqn2.8}
\sup_{E \in \RR} \sup_{a >0} |a^{1-\alpha}  \int Im(F^\omega(E+ia)) ~ d\mu(q^\omega)|  < d_\mu^{\alpha, \infty}.
\end{equation}
Let $E \in \RR$ and $a>0$, then we have using the well known rank one perturbation
formula for the resolvents (see Lemma 3.1.1 Demuth-Krishna \cite{dk} for example) that
\begin{equation}\label{eqn2.9}
Im(F^\omega(E+ia)) =  Im(\frac{1}{q^\omega + F_0^\omega(E+ia)^{-1}}) 
\end{equation}
and the assumption on $A^\omega$ and $q^\omega$ imply that the random variables
$F_0(E+ia)$ and $q$ are independent.
Therefore we have
\begin{equation}\label{eqn2.10}
\begin{split}
Im(\int F^\omega(E+ia) ~ d\mu(q^\omega))  &=  Im(\int \frac{1}{q^\omega + F_0^\omega(E+ia)^{-1}}) ~ d\mu(q^\omega) ) 
\end{split}
\end{equation}
Thus using equations (\ref{eqn2.71}) and (\ref{eqn2.11})
\begin{equation}\label{eqn2.11}
Im(\int \frac{1}{x - E - ia} ~ d\nu_{_{A^\omega}}(x)) = 
 Im(\int \frac{1}{q^\omega + F_0^\omega(E+ia)^{-1}} ~ d\mu(q^\omega) ).
\end{equation}
Now $F^\omega_0(E+ia) = \int \frac{1}{x-E-ia} d\sigma^\omega(x)$ for some probability measure
$\sigma^\omega$ ( in fact $\sigma^\omega(\cdot) = \langle \phi, E_{A^\omega}(\cdot) \phi\rangle$), which is independent of $q^\omega$ by assumption, so fixing it we have
\begin{equation}\label{eqn2.12}
Im(\int \frac{1}{x - E - ia} ~ d\nu_{_{A^\omega}}(x)) = 
Im(\int \frac{1}{x + F_0^\omega(E+ia)^{-1}} ~ d\mu(x)).
\end{equation}
From this we see that 
\begin{equation}\label{eqn2.13}
a^{1-\alpha}Im(\int \frac{1}{x - E - ia} ~ d\nu_{_{A^\omega}}(x)) = 
 a^{1-\alpha} Im(\int \frac{1}{x + F_0^\omega(E+ia)^{-1}} ~ d\mu(x) ).
\end{equation}
The integral in the expectation on the right hand side is uniformly bounded
by $2\pi ~  d_\mu^{\alpha, \infty}$ by Lemma \ref{lem2.3}). 
since $\mu$ is  uniformly $\alpha$-H\"older
continuous. On the other hand using the inequality (\ref{eqn002}), noting that
$\beta = 1$ there, in the case when $\psi(x) = \frac{1}{1+x^2}$, the left hand side has the lower bound, 
\begin{equation*}
\begin{split}
a^{1-\alpha} Im(\int \frac{1}{x - E - ia} ~ d\nu_{_{A^\omega}}(x)) &=
\frac{1}{a^\alpha}(\psi_a *\mu)(E) \\ &\geq 2^{\alpha -1} 
\frac{\nu_{_{A^\omega}}( (E-a, E+a) )}{(2a)^\alpha}. 
\end{split}
\end{equation*}
Therefore we get 
$$
\frac{\nu_{_{A^\omega}}((x-a, x+a))}{(2a)^\alpha} \leq 2^{1-\alpha} 2 \pi ~ d_\mu^{\alpha, \infty},
$$
which gives the required bound by taking sup over $a$ and $x$ on the left
hand side.  \qed 

{\noindent \bf Proof of theorem \ref{thm1.2}:}  The proof of this theorem 
proceeds on the same lines of that of theorem \ref{thm1.1} since $q^\omega(n)$ 
is distributed according to the probability measure $\mu_{a_n^{-1}}$, using the
comments after Remark \ref{rem1.1}.  Using
this fact and the equation (\ref{eqn1.1}) we obtain the bound
$$
d_{\nu_n}^{\alpha, \infty} \leq |a_n|^{-\alpha} d_\mu^{\alpha, \infty}.
$$
This estimate gives the bound
$$
d_{\nu}^{\alpha, \infty} \leq \sum \beta_n |a_n|^{-\alpha} d_\mu^{\alpha, \infty},
$$
from which the stated uniform $\alpha$-H\"older continuity of $\nu$ follows. \qed

{\noindent \bf Acknowledgement:}  We thank Werner Kirsch and Peter Stollmann for
discussions that improved the presentation of this work.

\end{document}